\documentstyle[osa,manuscript]{revtex}

\title{Gyration radius of a circular polymer 
under a topological constraint with  excluded volume 
}

\author{
Miyuki K. Shimamura and Tetsuo Deguchi}

\address{
Department of Physics, Faculty of
Science 
and Graduate School of Humanities and Sciences, 
Ochanomizu University 
2-1-1 Ohtsuka, Bunkyo-ku, Tokyo 112-8610, Japan
}

\begin{document}

\maketitle

\begin{abstract}
It is nontrivial whether the average size of 
a ring polymer should become smaller or larger
under a topological constraint. 
 Making use of some knot invariants,
we evaluate numerically the mean square radius of gyration  
for ring polymers having a fixed knot type,
 where the ring polymers are given by self-avoiding
polygons consisting of freely-jointed  hard cylinders.
We obtain plots  of the gyration radius versus
the number of polygonal nodes
for the trivial, trefoil and figure-eight  knots.
We  discuss  possible asymptotic behaviors
of the gyration radius under the topological constraint.
In the asymptotic limit, the size of a ring polymer with a given knot
is larger than that of no topological constraint 
when the polymer is thin, and the effective expansion 
becomes weak when the polymer is thick enough.
\end{abstract}
PACS number(s): 36.20.-r, 61.41.+e, 05.40.Fb

\newpage 
%%%%%%%% introduction %%%%%%%%%%%%%
\section{Introduction}

The effect of a topological constraint should be nontrivial
on physical quantities of a ring polymer such as the size of the ring polymer.
Once a ring polymer is formed, its topological state,
which is given by a knot, is fixed. However, it has  not been established
how to formulate the topological constraint on the ring polymer
in terms of analytic methods.
On the other hand, several numerical simulations have been performed,
investigating some statistical  properties of ring polymers
under topological constraints
\cite{Vologodskii1,desCloizeaux,Michels,LeBret,Chen,Klenin,Janse,Koniaris,DeguchiJKTR,Deguchi95,DeguchiRevE,Orlandini,Janse1991}.
 Through the simulations, it has  been found that
a topological constraint may severely restrict the available degrees
of freedom in the configuration space of a ring polymer,
and can be significant in its physical properties.

\par
Recently, DNA knots are synthesized in experiments,
and then separated into various knot types
by the agarose-gel-electrophoresis technique
\cite{Rybenkov,Shaw,StasiakNATURE,StasiakJMB}.
Under the electric field,  the charged macromolecules
move through the  network of the gel,
 and the migration rates should depend on their sizes, shapes and  charges.
It is  remarkable  that the electrophoretic mobility of a
circular DNA  depends also on its knot type.
It is observed  that the more complicated the DNA knot is,
the higher its mobility.
The fact could be related to a common belief
 that the gyration radius of a knotted DNA should depend on its knot type.

\par
In this  paper, we  discuss the mean square radius of gyration
of circular polymers having a fixed knot type.
We consider the question  how the size of a ring polymer
should  depend on  the length $N$  under the topological constraint.
Here the length $N$ corresponds to the number of polygonal nodes,
when we model the ring polymers  by some self-avoiding polygons.
We find that the effect of the topological constraint  is not trivial.
 In fact, the mean square radius of gyration for ring polymers
under the topological constraint can be smaller or larger than
that of the ring polymers under no topological constraint,
as we shall see later.
We study the question through numerical simulations,
 making use of some knot invariants.
We construct an ensemble of ring polymers consisting
of freely jointed hard cylinders,
and then  evaluate the mean square radius of the gyration
for the ring polymers that have the given knot type.

%%%%%%%%%%%%%% method I: cylinder model %%%%%%%%%%%%%%%%%%%%%%%%
\section{Cylindrical self-avoiding polygons}

Let us explain the model of ring polymers consisting of freely jointed hard
cylinders \cite{Klenin,PLA}.
The segments  are given by hard  cylinders  with radius $r$.
They are ``hard'' in the sense that
there is no overlap allowed between any pair of non-adjacent
cylindrical segments, while adjacent segments can overlap:
there is no constraint on any  pair of adjacent cylinders.
The model was first introduced in Ref. \cite{Klenin}
with the Monte-Carlo algorithm using the ``hedgehog'' configurations
of polygons.
Recently, another method was introduced
for constructing the cylindrical self-avoiding polygons \cite{PLA}.
It is based on the algorithm of ring-dimerization  \cite{Chen} for
the rod-bead model of self-avoiding polygons.
 We call the new algorithm
the cylindrical ring-dimerization method, or  the dimerization method,
 for short.
All the cylindrical self-avoiding polygons in this paper
are constructed by the dimerizaion method.
We note that the  algorithm of chain dimerization
 is quite effective for constructing long self-avoiding
walks on off-lattice \cite{Madras}.

\par
The most useful property   of the model of cylindrical self-avoiding
polygons is that it has the two parameters: the cylinder radius $r$ and
the number $N$ of segments.
The two parameters are important in  the theoretical explanation of
the experimental data of DNA knots \cite{Shaw}:
the cylinder radius corresponds to the effective radius
of negatively charged DNAs surrounded by the clouds of counter ions.
Thus, the  model is closely related to the wormlike chain model
for polyelectrolytes in electrolyte solutions.
Furthermore, it has been shown that the cylinder radius $r$
plays an important role
in the random knotting probability \cite{PLA}.

\section{Gyration radius of  cylindrical self-avoiding polygons}

Now let us consider the mean square radius of gyration $R^2$
for the cylindrical self-avoiding polygons.
It is defined by
\begin{equation}
R^2\equiv \frac{1}{2N^2}\sum_{n,m=1}^N<(\vec{R}_n-\vec{R}_m)^2>.
\end{equation}
Here $\vec{R}_n$ is the position vector of the $n$-th monomer and $N$
 is the number of nodes in the cylindrical self-avoiding polygon.
In numerical simulations, the statistical average $< \cdot >$ is given
by the average over $M$ samples of polygons.
 We take $M=10^4$ in the paper:
we construct $10^4$ self-avoiding polygons of $N$ segments 
of cylinders with radius $r$ by the dimerization method.

\par
We have calculated numerical estimates of the
the mean square radius of gyration  $ R^2 $  for the cylindrical
self-avoiding polygons with the different values of radius $r$
for several different  numbers  $N$ of polygonal nodes  upto 1000.
We consider eighteen numbers from 20 to 1000  for the number  $N$ of
polygonal nodes,
and  fifteen different values from 0.0 to 0.07 for the cylinder radius $r$.
Here we note that the gyration radii for the rod-bead polygons  
have been  estimated in Refs. \cite{Koniaris,turuD}.

%%%%%%%%%%%%%%% method II: random knotting  %%%%%%%%%%%%%
\section{Method for selecting  polygons with the same given knot}

\par Let us describe the method for selecting
such polygons that have a given knot type in our simulation.
 For a given knot $K$,
we enumerate the number $M_K$ of such polygons out of the $M$ polygons
that have the same set of values of some knot invariants for the knot type
$K$.
We employ two knot invariants, the determinant $\Delta_K(-1)$ of knot and
the Vassiliev-type invariant $v_2(K)$ of the second degree,
as the tool for detecting the knot type of a given polygon
\cite{DeguchiPLA,Polyak}.

\par
The number $M_K$ depends not only
on the knot type but also on the step number $N$ and cylinder radius $r$.
Let us recall the  knotting probability for cylindrical 
self-avoiding polygons consisting of cylinders with the radius
$r$ \cite{PLA}. 
We denote by $P_{triv}(N,r)$ the probability
 of a cylindrical self-avoiding polygon of $N$ nodes 
with cylinder radius $r$ being a trivial knot.   
Then, it is given by
\begin{equation}
P_{triv}(N,r) = \exp(-N/N_{c}(r)).
 \label{pro}
\end{equation}
Here  $N_c(r)$ is called the characteristic length of
random knotting and can be approximated  by an exponential function of $r$:
$N_c(r) = N_c(0) \exp( \gamma r)$, where $\gamma$ is a constant \cite{PLA}.

\section{Gyration radius of ring polymers under a topological constraint}
%%%%%%%%%%%%%%%%%%%%% ring polymer of a specific knot %%%%%%%%%%%%%

Let us discuss the mean square radius of  gyration
for cylindrical self-avoiding polygons with a fixed knot type.
We evaluate  the expected value of the gyration radius for a given
knot $K$ by the  mean value of the square  radii of gyration
for  the $M_K$ polygons with the knot $K$.
Then, the mean square radius of gyration  $R_K^2$
for the knot type $K$ is given by
\begin{equation}
R_K^2=\frac{1}{M_K} \sum_{i=1}^{M_K} R_{K,i}^2,
\end{equation}
where $R_{K,i}^2$ denotes  the gyration radius of the $i$-th 
cylindrical self-avoiding polygon that has the knot type $K$,
 in the set of $M$ polygons.
For the rod-bead model of self-avoiding polygons,
$R_K^2$ has been evaluated for some different knots \cite{turuD}.

\par
Let us discuss our data of  numerical estimates of $R_K^2$.
For trivial, trefoil and  figure-eight knots,
we find  that the mean square radius of gyration
increases monotonically with respect to  the number  $N$ of polygonal nodes.
However, the ratio $R_K^2/R^2$ is not constant with respect to $N$.
The graphs of the ratio $R_K^2/R^2$ against the step number $N$
are plotted in Fig. 1 for  trivial and trefoil knots.
Here the cylinder radius is given by 0.005, i.e.,
the diameter is given by 0.01.
Here we recall that $R^2$ denotes the mean square radius of gyration
for the self-avoiding polygons with all possible knot types.
In terms of $R_K^2$,  $R^2$ is given by
the following: $R^2 = \sum_{K} M_K R_K^2/M$.
We recall that the number $M_K$ gives different values
for different step numbers $N$ or different  knots $K$.
When the knot $K$  is complicated,  $M_K$ can be  very small and
it may give a poor statistics to $R_K^2$.

\par
Let us discuss the plots of the trivial knot shown in Fig. 1.
The ratio $R^2_{triv}/R^2$ at $N=21$ is almost given by 1.0.
This is consistent with the fact that
 trivial knots are dominant when $N$ is small. Here we recall that
 the  probability of being a trivial knot is given by eq. (\ref{pro}).
On the other hand, the ratio $R^2_{triv}/R^2$ increases
with respect to the step number $N$. Thus, the size of the ring polymer
enlarges under the topological constraint of being  a trivial knot.
The topological constraint gives an effective swelling effect, in this case.

\par
Let us consider the case of  trefoil knot.
 The ratio $R^2_{tre}/R^2$ is not always larger than 1.0. 
In fact, it is  smaller than 1.0 when $N < 200$.
The ratio $R^2_{tre}/R^2$ is given by about 0.7 when $N=21$.
Thus, the topological constraint gives an effective
shrinking effect on the ring polymer.
When $N > $ 300 or 400, the ratio $R^2_{tre}/R^2$ 
becomes larger than 1.0 for $r=$ 0.005.
Thus, the topological constraint makes the ring polymer enlarge for large $N$.
However, when $r> 0.03$, the ratio $R^2_{tre}/R^2$ becomes smaller than 1.0 
even for $N=1000$, as shown in Fig. 2.

\par 
In Figs. 1 and 2,  we see that the difference among the ratios of the
gyration radii for trivial,
trefoil and figure-eight knots 
is much more clear  in the small-$N$ region than in the large-$N$ region.
The dependence of the gyration radius
of the ring polymer  on its  knot type could be more significant
for small $N$ than for large $N$. 
The effective shrinking of a thick and finite ring-polymer with a nontrivial knot 
might be associated with  the concept of ideal knots or tight knots
\cite{idealknot,Grosberg}.

\par
The fitting curves in Figs. 1 and 2 are given by
\begin{equation}
{\frac {R_K^2} {R^2}} =  \gamma_K   
\left(1 - \delta_K \exp(- \eta_K N ) \right) \, . 
\label{formula}
\end{equation}
Here, the constants $\gamma_K$, $\delta_K$ and $\eta_K$ are fitting parameters. 
We see that the fitting curves in Figs. 1 and 2 are consistent with all 
the numerical estimates of $R_K^2$ in the range from $N=21$ to 1000.  
Thus, the formula (\ref{formula}) effectively 
describes the finite-size behaviour of $R_K^2$,  
although there is no apriori reason for assuming it.  
For instance,  the formula (\ref{formula}) can be 
not appropriate as an asymptotic expansion. 
In \S 6, we shall introduce another formula to discuss 
 the possible asymptotic behaviors of $R_K^2$.

\section{Asymptotic behaviors of $R_K^2$ }

Let us discuss  possible asymptotic behaviors 
of the gyration radius of the ring polymer under a topological constraint.
We may assume that when $N$ is very large, $R_K^2$ can be approximated by 
\begin{equation}
R_K^2 = A_K N^{2 \nu_K} \left( 1+ B_K N ^{-\Delta_K} + O(1/N) \right) \, .   
\label{asymp} 
\end{equation} 
The expansion is consistent with  renormalization group arguments, 
and hence it should be valid for the case of asymptotically large $N$. 
The exponent $\nu_K$ and the amplitude  $A_K$ can be evaluated by 
applying the formula (\ref{asymp}) to the numerical data of $R_K^2$ 
for large values of $N$.  Here we note that 
the expansion (\ref{asymp}) is not effective for small $N$. 
In fact, when $N \le  200$,   it does not give any good fitting curves 
to the data in Figs. 1 and 2.

\par 
Let us now discuss the exponent $\nu_K$. 
We have analyzed the plots of $R^2_K/R^2$ given in Fig. 1 
for trivial and trefoil knots  by the least-sqaure method with 
respect to the following formula:   
${R_K^2}/{R^2}  = ( {A_K}/A) N^{\nu_K-\nu} 
\left( 1+ (B_K-B) N ^{-\Delta} + O(1/N) \right)$. Here we have assumed 
$\Delta_K = \Delta = 0.5$, which is consistent with the scaling expansion.   
Then, we obtain the following  estimates: for trivial knot, 
 $\nu_K-\nu=0.005 \pm 0.103$, $A_{K}/A=1.18 \pm 0.99$, $B_{K}-B = - 1.7 \pm 4.0$; 
for trefoil knot, $\nu_{K}-\nu=0.009 \pm 0.090$, $A_{K}/A=1.18 \pm 0.85$, 
$B_{K}-B = - 3.5 \pm 3.0$. To each of the knots, 
we have applied the formula (\ref{asymp}) to the eight points 
with $N \ge 300$ in Fig. 1.  
The  $\chi^2$ values are given by 4.3 and 6.2 for trivial and trefoil knots, respectively.     
 For other values of radius $r$, we have similar results.   
Thus, for trivial and trefoil knots,  the exponent $\nu_K$ should agree with 
the exponent $\nu$ of the mean square radius of gyration $R^2$, 
within the error bars.

\par   
There are also other evidences supporting $\nu_K=\nu$, i.e., $\nu_K= 0.588$, 
 for trivial and trefoil knots. In fact, the plots of the ratio $R_{K}^2/R^2$ 
versus the number $N$  in Figs. 1 and 2  are likely to approach some horizontal lines
at some large $N$. 
 It is also the case with  
 some other values of cylinder radius $r$. 
 It is  clear particularly for  trivial knot. 
 Thus, at least for trivial knot, we can easily conclude that  
  the exponent $\nu_K$  should coincide with the exponent $\nu$. 
 We  note that even the  fitting formula (\ref{formula}) is  consistent with 
 the coincidence of the exponents: $\nu_K=\nu$. Thus, 
 all the numerical results obtained  in the paper 
suggest $\nu_K=\nu$ for trivial and trefoil knots. 
 It  is also consistent with the observation for the self-avoiding  polygon on
a lattice in Refs. \cite{Orlandini,Janse1991}
that the exponent $\nu_K$ should be  independent of the knot type.

%%%%%%%%%%%%%%%%%%%%%%  DATA ANALYSIS %%%%%%%%%%%%%%%%%%%%

\par 
Let us  now consider the amplitude $A_K$ of the asymptotic expansion 
(\ref{asymp}). In Fig. 3,  
the numerical plots of the ratio $A_K/A$ versus the radius $r$ 
are shown  for trivial and trefoil knots.  The ratio is  evaluated  by the following 
formula:  
${R_K^2}/{R^2}  = ({A_K}/A) \left( 1+ (B_K-B) N ^{-\Delta} + O(1/N) \right) $
Here, we have assumed  $\nu_K = \nu$ and $\Delta_K=\Delta$=0.5.      
 Furtheremore, we have applied it only to the data with $N \ge 300$.

\par 
When $r$ is small, the ratio $A_K/A$ in Fig. 3 is larger than 1.0 
for both  trivial and trefoil knots.
 It is  remarkable that the asymptotic ratio $A_K/A$ can be larger than 1.0 
when the ring polymer is thin enough.  We can easily confirm it also in Fig. 1,  
observing the increasing behavior of the plots of $R^2_{triv}/R^2$ versus $N$. 
 Thus,  the topological constraint gives 
an effective expansion to the ring polymer with small radius $r$.

\par 
Interestingly, we see in Fig. 3 
that the ratio $A_K/A$ decreases monotonically
with respect to the  cylinder radius $r$. 
One might expect that  the ring polymer with larger excluded-volume 
should become  much larger.
In reality,  however, for the ring polymer under the topological constraint,  
the ratio $A_K/A$ decreases  
if the excluded-volume parameter becomes large. 
It is as if the excluded volume effect could make weaken 
 the effective expansion derived from the topological constraint. 
We notice in Fig. 3 that 
  $A_K/A$ might become close to the value 1.0 when $r$ is thick enough. 
Due to the poor statistics, we can not determine whether it really does or not.   
However, if it does, then it is consistent with the interpretation on the lattice model 
of  Refs. \cite{Orlandini,Janse1991} that $A_K$ should be independent 
of knot type.

\par 
 It is  quite nontrivial that  $A_K/A$ which is valid in the asymptotic expansion 
can be larger than 1.0, and the ratio decreases with respect to the radius $r$. 
Let us give one  possible explanation for it. 
First, we note that $R^2$ is given 
by the average of $R_K^2$ over all possible knots: $R^2= \sum_K R^2_K P_K(N)$.  
The fact: $A_K/A > 1.0$ suggests that there are large number of 
knots smaller than the knot $K$. 
Second, we recall the finite-$N$ behavior of $R_{tre}^2$: 
 In Figs. 1 and 2 we see that the ratio $R_{tre}^2/R^2$  is much smaller 
than 1.0 when $N$ is small, while it increases with respect to $N$ 
and finally becomes constant. 
We consider that when $N$ is small, the size of trefoil knot 
is small due to a finite-size effect, while 
when $N$ is  asymptotically large  then   
 $R_{tre}^2/R^2$ becomes almost constant.  
 For a given knot $K_1$, if we take $N$ large enough, then 
the ratio $R_{K_1}^2/R^2$ becomes constant with respect to $N$, 
while the majority of knots possible in $N$-noded polygons 
 should be much more complex than the knot $K_1$ and their sizes should be much smaller 
 than that of the knot $K_1$. Thus, 
 $R^2$ can be smaller than 
 $R_{K_1}^2$ and the ratio $A_{K_1}/A$ can be larger than the value 1.0.

\par 
 We remark that $R^2_{triv}/R^2$ is always larger than 1.0 
 both for finite $N$ and asymptotically large $N$.   
 For the finite-$N$ case, it is clear from Figs. 1 and 2 that $R_{triv}^2/R^2 > 1$. 
 For the asmptotic case, it is suggested  from Fig. 3  
that $A_{triv}/A$ should be larger than 1.0 for some small values of 
radius $r$. Thus, the property: $R^2_{triv}/R^2 > 1 $ 
 should persist in the asymptotic limit,  
as far as the data analysis is concerned.

\par
 Summarizing the numerical results on the asymptotic behaviors,
we conclude that the topological constraint on a ring polymer gives
an effective expansion  when the radius $r$ is small and also that 
 the expanding effect becomes weaker when the radius $r$ becomes larger,
and it may vanish when the radius $r$ is large enough. 
The results  suggest that there should be a new `phase transition',
where it is controlled by the excluded-volume parameter
whether the topological constraint gives  an effective expansion to the ring polymer or not.

%%%%%%%%%%%%%%%%%%%%%%%%%%%%%%%%%%%%%%%%%%%%%%%%%%%%%%%%%%%%

\newpage

%%%%%%%%%%%%%%%%%%%%%%%%%%%%%%%%%%%%%%%%%%%%%%%%%%%%%%%%%%%

\begin{figure}
\caption{The ratio $R^2_K/R^2$ versus the number $N$ of polygonal nodes of the
cylindrical self-avoiding polygons.
Numerical estimates of $R^2_{triv}/R^2$ for $r=$ 0.005 are shown by black
circles and those of $R^2_{tre}/R^2$ for $r=0.005$ by black triangles.
In the inset, the enlarged figure shows the numerical estimates 
of  $R^2_K/R^2$  from $N=$20 to 100 for the cases of trivial, 
trefoil and figure-eight knots. }
\end{figure}

\begin{figure}
\caption{The ratio $R^2_K/R^2$ versus the number $N$ of polygonal nodes of the
cylindrical self-avoiding polygons.
Numerical estimates of $R^2_{triv}/R^2$ for $r=0.04$   are shown by black
circles and those of $R^2_{tre}/R^2$ for $r=0.04$ by black triangles.
The fitting parameters are given by the follwing:  for trivial knot, 
 $\nu_{K}-\nu=-0.0004 \pm 0.068$, $A_{K}/A=1.04 \pm 0.59$, $B_{K}-B = - 0.3 \pm 3.1$, 
and $\chi^2=1.1$; for trefoil knot, $\nu_{K}-\nu=0.006 \pm 0.101$, $A_{K}/A=1.01 \pm 0.84$, 
$B_{K}-B = - 2.0 \pm 4.0$, and $\chi^2=3.2$.    
In the inset, the enlarged figure shows the numerical estimates 
of  $R^2_K/R^2$  from $N=$20 to 100 for the cases of trivial, 
trefoil and figure-eight knots. }
\end{figure}

\begin{figure} 
\caption{The ratio $A_K/A$ versus the cylinder radius $r$ for the cylindrical 
self-avoiding polygons.
The values of $A_K/A$ for the trivial knot and the
trefoil knot are shown by black circles and black triangles, respectively.
Each of the black triangles are slightly shifted rightward 
by one ten-th of the value of radius $r$, for graphical convenience. }
\end{figure}


\begin{thebibliography}{[99]}

\bibitem{Vologodskii1} A.V. Vologodskii, A.V. Lukashin, M.D.
Frank-Kamenetskii, and V.V. Anshelevich,
Sov. Phys. JETP {\bf 39},1059 (1974).

\bibitem{desCloizeaux} J. des Cloizeaux and M.L. Mehta,
J. Phys. (Paris) $\bf{40}$, 665 (1979).

\bibitem{Michels} J.P.J. Michels and F.W. Wiegel,
Phys. Lett. A $\bf{90}$, 381 (1982).

\bibitem{LeBret} M. Le Bret, Biopolymers $\bf{19}$, 619 (1980).

\bibitem{Chen} Y.D. Chen, J. Chem. Phys. $\bf{74}$, 2034 (1981) ; J. Chem.
Phys. $\bf{75}$, 2447 (1981); J. Chem. Phys. $\bf{75}$, 5160 (1981).

\bibitem{Klenin} K.V. Klenin, A.V. Vologodskii, V.V. Anshelevich, A.M.
Dykhne and M.D. Frank-Kamenetskii, J. Biomol. Struct. Dyn. $\bf{5}$, 
1173(1988).

\bibitem{Janse} E.J. Janse van Rensburg and S.G. Whittington, J.Phys. A
$\bf{23}$, 3573 (1990).

\bibitem{Koniaris} K. Koniaris and M. Muthukumar,
Phys. Rev. Lett. $\bf{66}$, 2211 (1991).


\bibitem{DeguchiJKTR} T. Deguchi and K. Tsurusaki,
J. Knot Theory and Its Ramifications $\bf{3}$, 321 (1994).


\bibitem{Deguchi95} T. Deguchi and K. Tsurusaki,
in {\it Geometry and Physics}, Lect. Notes in Pure and
Applied Math. Series/184, ed. by
J.E. Andersen, J. Dupont, H. Pedersen, and A. Swann,
(Marcel Dekker Inc., Basel Switzerland, 1997),  pp. 557-565.

\bibitem{DeguchiRevE} T. Deguchi and K. Tsurusaki,
Phys. Rev. E $\bf{55}$, 6245 (1997).

\bibitem{Orlandini} E. Orlandini, M.C. Tesi, E.J. Janse van Rensburg and
S.G. Whittington,
J. Phys. A: Math. Gen. {\bf 31}, 5953 (1998).

\bibitem{Janse1991} E.J. Janse van Rensburg and S.G. Whittington,
 J.Phys. A $\bf{24}$, 3935 (1991).

\bibitem{Rybenkov} V.V. Rybenkov, N.R. Cozzarelli and A.V. Vologodskii,
Proc. Natl. Acad. Sci. USA {\bf 90}, 5307 (1993).


\bibitem{Shaw}
S.Y. Shaw and J.C. Wang,
Science {\bf 260}, 533 (1993).


\bibitem{StasiakNATURE} A. Stasiak, V. Katritch, J. Bednar, D. Michoud and
J. Dubochet, Nature {\bf 384}, 122 (1996).


\bibitem{StasiakJMB} A.V. Vologodskii, N. J. Crisona, B. L. P. Pieranski, V.
Katritch,
J. Dubochet and A. Stasiak, J. Mol. Biol. {\bf 278}, 1 (1998).


\bibitem{PLA} M.K. Shimamura and T. Deguchi,
Phys. Lett. A  {\bf 274}, 184 (2000) 184; see also, 
M.K. Shimamura and T. Deguchi, to appear in J. Phys. Soc. Jpn. {\bf 70} No. 6 (2001). 


\bibitem{Madras} N. Madras and G. Slade, {\it The Self-Avoiding Walk},
(Birkh{\"a}user, Boston, 1993), \S 9.3.2.


\bibitem{turuD} K. Tsurusaki, Thesis: Statistical Study of Random Knotting,
 (December 1994, Tokyo University)


\bibitem{DeguchiPLA}
T. Deguchi and K. Tsurusaki, Phys. Lett. A $\bf{174}$, 29 (1993); \\
See also, M. Wadati, T. Deguchi and Y. Akutsu, Phys. Reports {\bf 180}, 247
(1989) ; V.G. Turaev, Math. USSR Izvestiya {\bf 35}, 411 (1990).

\bibitem{Polyak} M. Polyak and O. Viro, Int. Math. Res. Not. No.11, 445 (1994).

\bibitem{idealknot}
 V. Katritch, J. Bednar, D. Michoud, R.G. Scharein,
J. Dubochet, and A. Stasiak,Nature {\bf 384}, 142 (1996).

\bibitem{Grosberg}
A. Yu. Grosberg, A. Feigel and Y. Rabin, Phys. Rev. E
{\bf 54}, 6618 (1996).


\end{thebibliography}
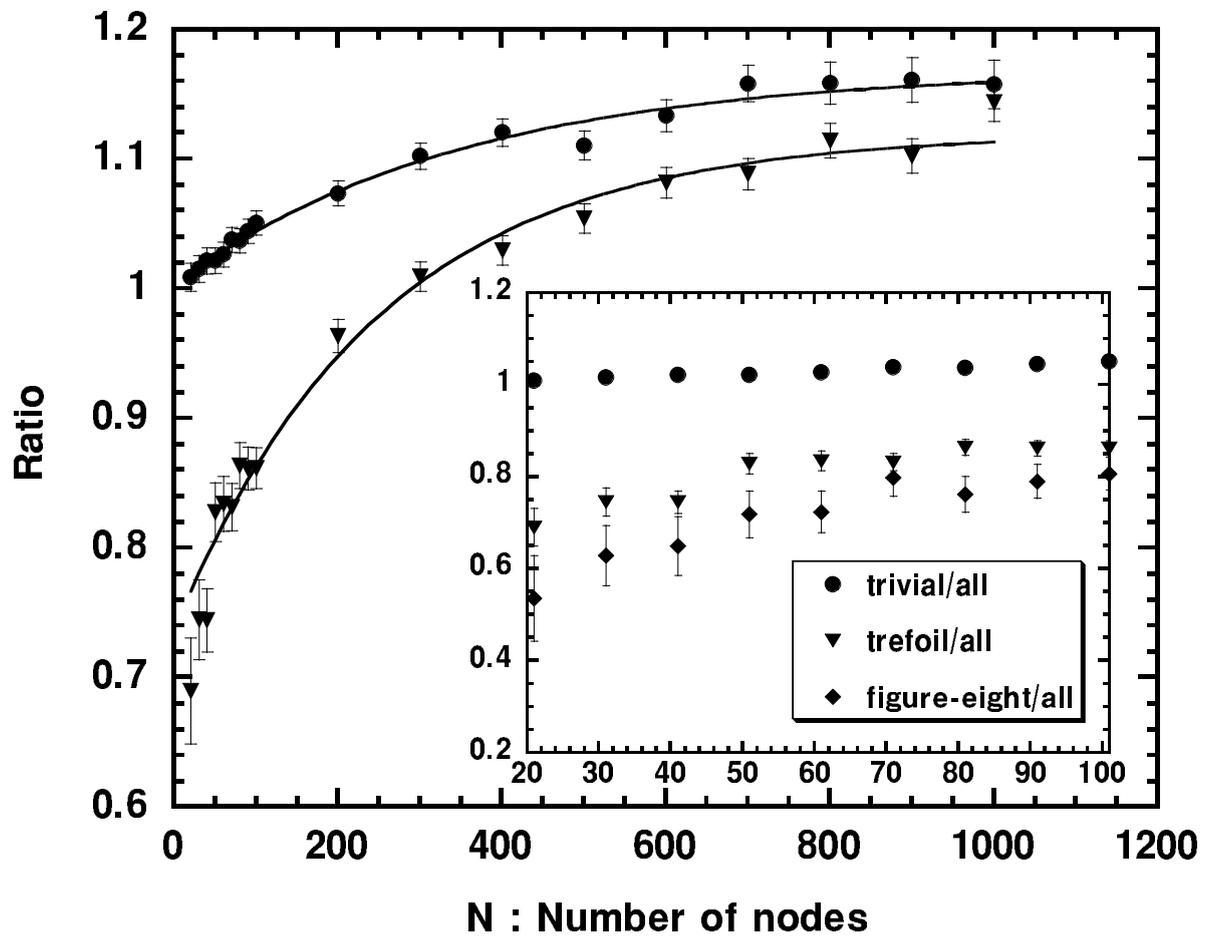
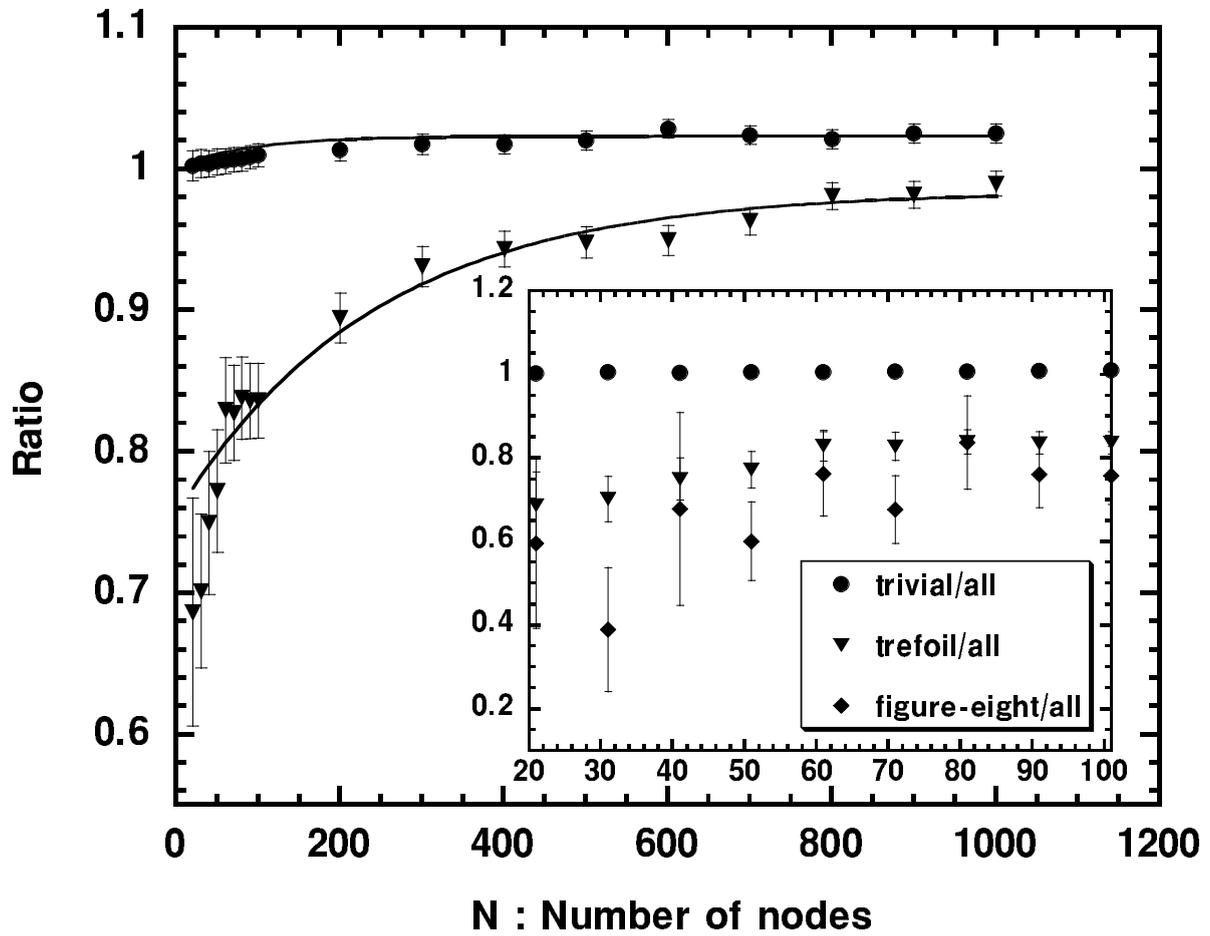
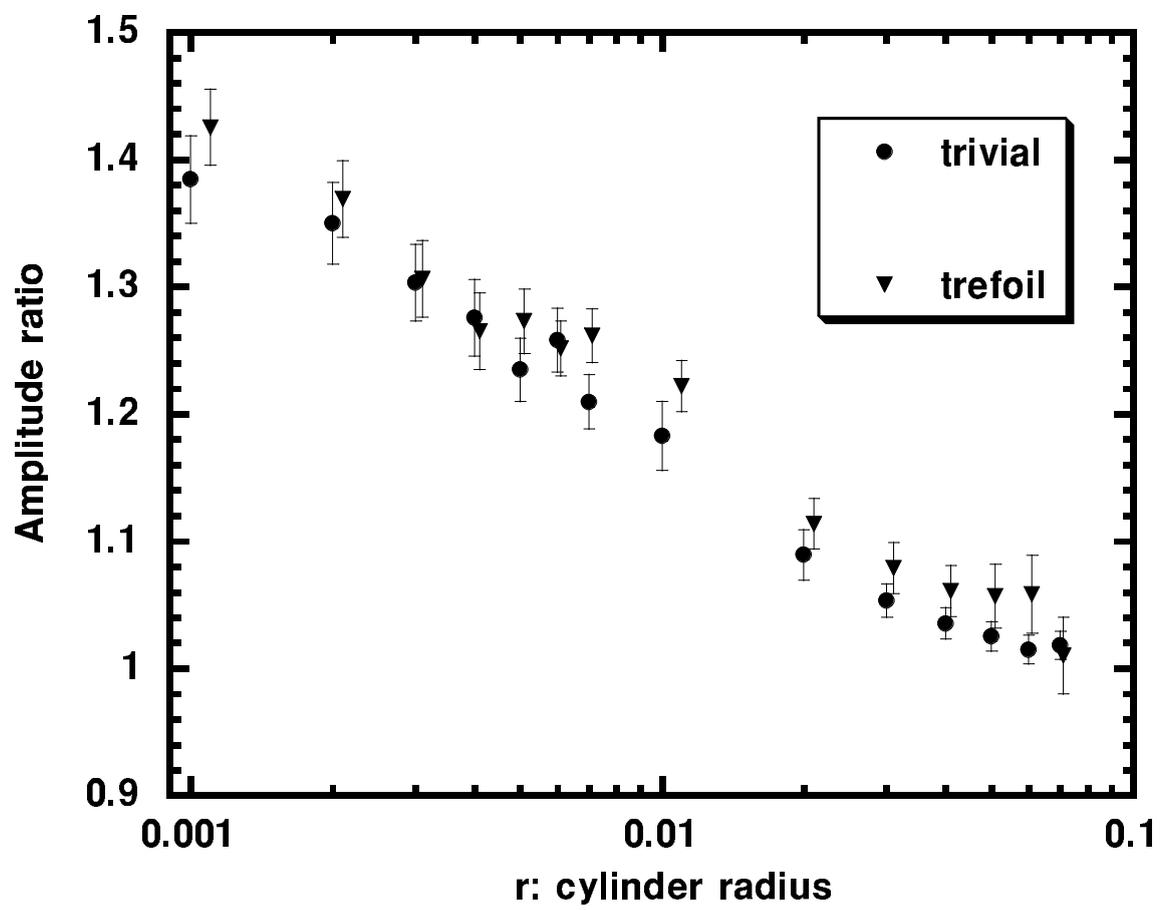
\end{document}